# Entombed

## An archaeological examination of an Atari 2600 game

John Aycock[a] and Tara Copplestone[b]

a Department of Computer Science, University of Calgary, 2500 University Drive NW, Calgary, Alberta, Canada

b Centre for Digital Heritage, University of York, Department of Archaeology, Heslington, UK

**Abstract** The act and experience of programming is, at its heart, a fundamentally human activity that results in the production of artifacts. When considering programming, therefore, it would be a glaring omission to not involve people who specialize in studying artifacts and the human activity that yields them: archaeologists.

Here we consider this with respect to computer games. We draw from the nascent archaeological subarea of archaeogaming to carry out a digital excavation of the code and techniques used in the implementation of Entombed, an Atari 2600 game released in 1982 by US Games. The player in this game is, appropriately, an archaeologist who must make their way through a zombie-infested maze. Maze generation is a fruitful area for comparative retrogame archaeology, because a number of early games on different platforms featured mazes, and their variety of approaches can be compared. The maze in Entombed is particularly interesting: it is shaped in part by the extensive real-time constraints of the Atari 2600 platform, and also had to be generated efficiently and use next to no memory. We reverse engineered key areas of the game's code to uncover its unusual maze-generation algorithm, which we have also built a reconstruction of, and analyzed the mysterious table that drives it. In addition, we discovered what appears to be a 35-year-old bug in the code, as well as direct evidence of code-reuse practices amongst game developers.

What further makes this game's development interesting is that, in an era where video games were typically solo projects, a total of five people were involved in various ways with Entombed. We piece together some of the backstory of the game's development and intoxicant-fueled design using interviews to complement our technical work.

Finally, we contextualize this example in archaeology and lay the groundwork for a broader interdisciplinary discussion about programming, one that includes both computer scientists and archaeologists.

## 1 Introduction

Donald Knuth famously characterized programming as an art through his series of *The Art of Computer Programming* volumes. This is telling. Art is a creative, human endeavor, and one that typically results in the production of (possibly ephemeral) artifacts. When viewed in that light, it seems clear that the study of programming overlaps with the artifact-based study of human activity: archaeology. Traditional archaeology primarily concerns itself with past human activity, and in our examination of programming we focus here on the recent past, and in particular the programming of old computer games.[1]

Play is also a human activity, of course, and it was both natural and inevitable that computers would be harnessed to play games. The early computers used for games were heavily constrained in multiple dimensions: memory, CPU power, speed, graphics, sound, storage capability. And yet, game programmers of the time cajoled these platforms into running their creations, laying the foundations for what has become a multi-billion-dollar industry whose products have cultural relevance.

How did these programmers accomplish their task under such constrained conditions? Necessity is the mother of invention, and indeed it has been observed in many fields that there is a link between constraint and creativity (e.g., [55]). Retrogame programmers used many clever programming tricks to make their creations work, and there is an opportunity to dig these out, study them, and preserve this knowledge, essentially conducting "retrogame archaeology."

We present this archaeology through a case study of the 1982 game Entombed, a game where the player – an archaeologist – had to traverse a maze populated by zombies. As described later, we used technical means to reverse engineer, analyze, and reconstruct the maze-generation algorithm, an algorithm that had to operate on the fly using next to no resources. We also were able to uncover a decades-old bug and, using that, demonstrate code reuse between games. By supplementing our technical work with interviews, we were able to learn more about the unusual design story behind this game and development practices that would not be apparent from the code alone.

Entombed was released for the Atari 2600 (Figure 1), a popular home game console in the early days of the game industry. Making its debut in 1977, the Atari 2600[2] was not the first home game console to appear, a distinction enjoyed by the Magnavox Odyssey of 1972. Nor was it even the first to sport game cartridges in a modern sense, having been beat to market by the Fairchild Channel F (1976) and the short-lived RCA Studio II (1977). However, the Atari 2600 did have a surprisingly lengthy lifespan: it was produced until 1992, well after its competitors' products had eclipsed the 2600's capabilities. Perhaps in part due to its longevity, in part being the first introduction many homes had to computers and video games [34], the Atari 2600 took on a degree of cultural importance; Montfort and Bogost observe that 'the generic term for a videogame system in the early 1980s was "an Atari"' [32, page 4]. Its simple one-button joystick is arguably iconic even today [32, 57], even appearing as a (largely) positive example of industrial design [29].[3] And, while not the sole cause, the Atari 2600 can be said to have had a substantial economic impact by virtue of its role in the 1983 North American video game crash [14].

[1] We use "computer games" in preference to "video games" because not all computer games require video.

[2] It was originally called the Atari Video Computer System (VCS), but it is perhaps better known under its later "Atari 2600" name, which we use throughout this paper for consistency.

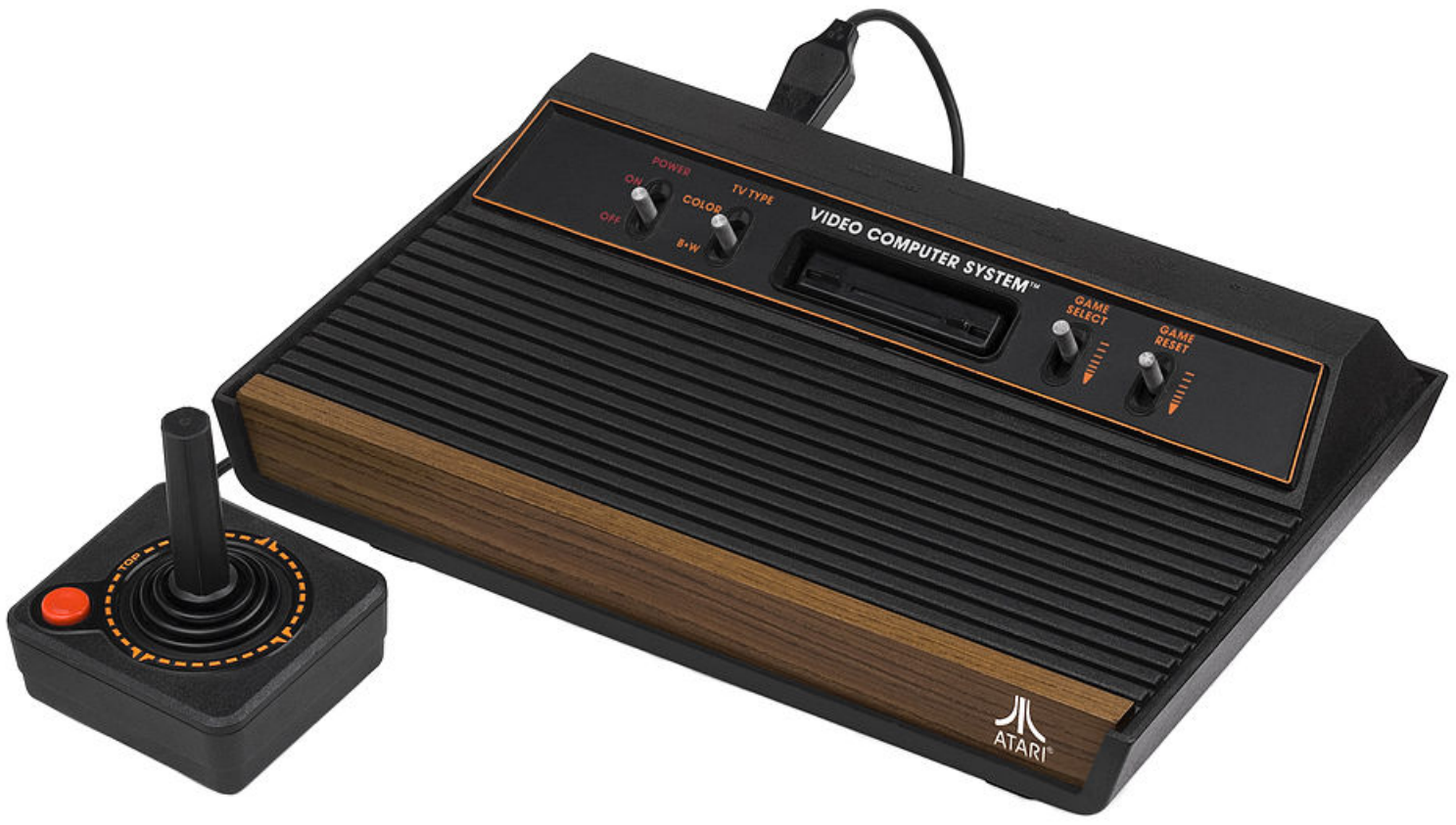

**Figure 1** The Atari 2600 (four-switch model) and one of its joysticks [3]

The contributions of this paper are fivefold.

- Adding to current understanding about the art of programming, by highlighting low-level programming techniques that are now less prominent, in the context of an extremely constrained platform (the Atari 2600) that was in widespread use for many years. This helps capture these techniques in the programming literature for modern and future audiences.
- Through our case study, increasing understanding of early procedural content generation by documenting Entombed's maze generation algorithm which, to our knowledge, is unique.
- Empirically demonstrating code reuse practices in early computer games through the "signature" of a buggy pseudo-random number generator.
- Helping to develop interdisciplinary methodology across disparate disciplines for studying software and the programming that created it. This allows a more diverse set of people, with different methodologies, to study programming.
- Specifically, working across disciplines with an archaeologist. Archaeology is a term easily purloined, but it is an established field older than computer science and includes more than working with old things. As a discipline, archaeology is well used to interpreting a wide range of evidence to understand human activity, and here we turn our attention to a digital artifact, Entombed. Already we can see through our case study that programmers' intention is lost, and evidence from the people involved (something which we will unfortunately not enjoy for much longer) is conflicting. As our software ages it will become increasingly important that appropriate methods for its study have been established.

[3] Foreshadowing Section 2, they note 'The greatest genius of the design may very well be the constraints it placed on game developers' [29, page 98].

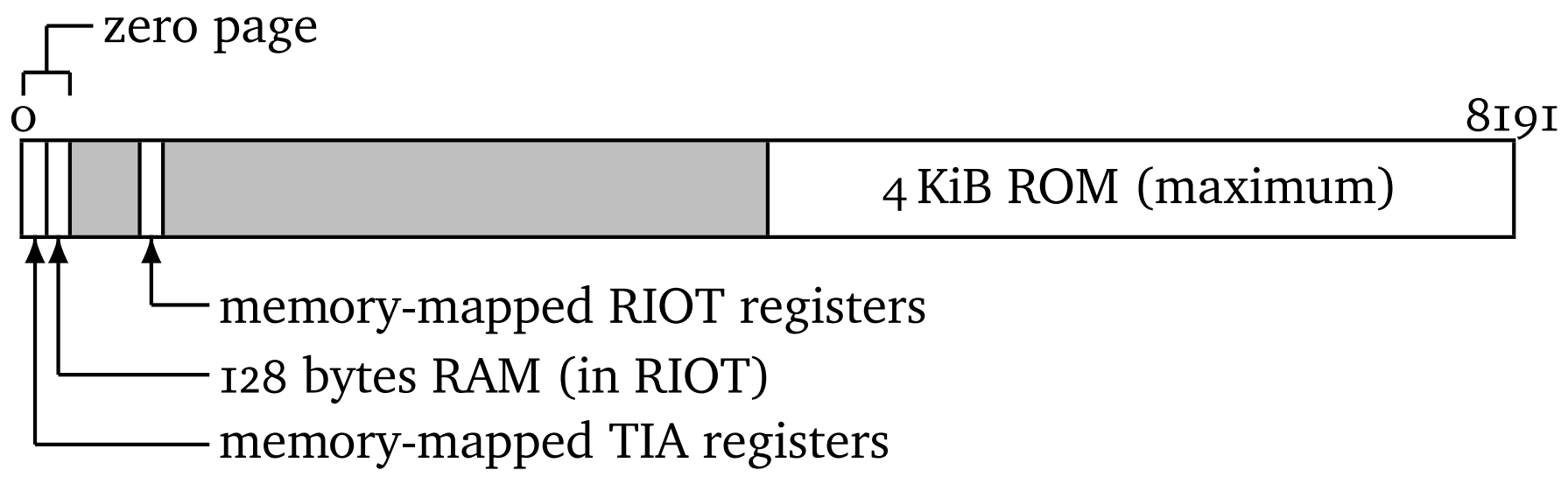

**Figure 2** Atari 2600 address space

From a technical point of view, the Atari 2600 had design decisions and constraints that made it an extremely challenging platform to program, which we examine in the next section. Section 3 presents our case study, Entombed, after which we contextualize our work (and programming more generally) with respect to archaeology in Section 4. Finally, Sections 5 and 6 discuss related work and our conclusions, respectively.

## 2 Programming the Atari 2600

In order to fully appreciate the constraints under which Atari 2600 programmers had to work, and to understand how Entombed was shaped by those constraints, we need to take a close look at the Atari 2600 platform from the programmer's point of view. We frame this discussion by beginning with a summary list of the 2600's programming constraints:

- Limited ROM (i.e., program) space.
- Very little RAM.
- No interrupts – polling only for I/O and timer.
- No video framebuffer.
- Real-time requirements, necessitating cycle counting.
- No operating system, BIOS, or pre-existing ROM routines.
- Possible lack of technical documents, and the need to reverse engineer the platform.

We now turn to the platform's design for context, to explain where these constraints came from. The Atari 2600's motherboard contained only three chips of note, a stark contrast to some of its competition – the Fairchild Channel F, for instance, had 40 chips [16]. Since the 2600's functionality was divided between those three chips, we use them to structure this tour of the platform. Unless stated otherwise, information in this section is drawn from [60]. Figure 2 shows the 2600's memory map[4] for reference throughout this section.

### 2.1 CPU

The CPU in the Atari 2600 was the 6507, a stripped-down version of the 6502, an 8-bit processor. The 6502 and variants thereof featured prominently in that era, and could be found powering game consoles (Atari 2600, Nintendo NES, PC Engine), home computers (Apple II, Commodore 64, BBC Micro, and many others), and even embedded inside peripherals (Commodore 1541 floppy disk drive). With only 28 pins rather than the 40 pins on the 6502, the 6507 was for all intents and purposes a full 6502 internally, but sacrificed some I/O lines to compensate for the smaller package.

The change with the largest effect between the 6502 and 6507 was the reduction in address bus lines. A full 6502 had 16 address lines for its 64 KiB address space; the 6507 exported only the low 13 lines, meaning that only 8 KiB of address space could be uniquely distinguished from outside the 6507, and the programmer saw extensive memory mirroring [10]. For example, the programmer would see the memory contents located at $f000 replicated at at $d000, $b000, $9000, and so on, because the three high bits of an address were unknown outside the 6507 and thus addresses could not be precisely decoded. These address space limitations also imply that game size was very limited, with later 2600 games transcending the 4 KiB ROM barrier only through "bank switching" that required additional hardware in the game cartridge.

Bank switching refers to the ability to have 1 of *N* different memory banks potentially available, all mapped to the same range in the address space; only one of them would be accessible at any given time. The programmer would select a memory bank via separate control lines [38] or, on systems like the Atari 2600 or Apple II, by accessing memory-mapped "soft switches" [4, 10].

Another 6507 tradeoff that affected how programs were structured was the lack of interrupt lines. Code could not be interrupt-driven, and all I/O had to be performed via polling.

### 2.2 RIOT

Some RAM was obviously required for the Atari 2600, and it was located inside an MOS 6532 – better known as a "RIOT" chip for its amalgamation of *R*AM, *I/O*, and *T*imer in one package [47]. This chip provides the entirety of the RAM on the system, a whopping 128 bytes of it.

This meager amount of memory had programmer-visible effects when combined with the 6507's architecture. The first 256 bytes of the 6507's address space, "zero page," was prime memory real estate: instructions referencing zero page were both shorter and faster. The 6507's fixed-position stack page was the *second* 256 bytes, as far as the 6507 was concerned. The 2600 quietly mapped the stack into zero page, however, meaning that the RIOT's 128 bytes did double duty. The programmer had to carefully balance stack usage with variable storage, as the Atari programmer's guide put it, 'hoping the two never meet' [60, page 21].

[4] For simplicity, the figure omits the memory mirroring we discuss later.

The I/O available through the RIOT for the programmer's polling pleasure included two four-position joysticks, and five of the six switches on the main console could also be polled, up to and including the switch labeled 'game reset.' Some I/O, like the joystick buttons and paddle controllers, was routed instead to the final chip, the TIA.

### 2.3 TIA

The TIA, or television interface adapter, was a custom Atari chip whose primary purpose was driving the video and audio signal of the TV that the Atari 2600 would be plugged in to.

The mention of RAM above did not mention any RAM used for the video framebuffer, and the reason for that was because the Atari 2600 didn't have one. In most systems, then and now, the programmer would conceptualize their game display in terms of high-level objects such as the player might see, placing them in memory that the hardware would automatically take care of rendering onscreen. Not so on the Atari 2600. A 2600 programmer had to structure their code to render the display on a *line by line* basis, and repeat that for *every* single video frame. The TIA's register values would be used to produce the output signal for any given line, but it was the programmer's responsibility to load those TIA registers with the correct values before the television's electron beam swept past, a process called 'racing the beam' [32].

The TIA's design, *sans* framebuffer, would have made sense to hardware engineers used to building video games using discrete logic, which involves a similar decomposition of screen images as a repeatedly redrawn 2D grid [5]. Effectively the design turned the Atari 2600 into a real-time system, where a failure by the programmer to meet the video timing constraints would result in the picture rolling. The time-critical display code in an Atari 2600 game was referred to as the "kernel," and programmers would typically count cycles in their assembly code to avoid overruns, leaving documentary evidence as assembly comments in the source code. Carol Shaw's River Raid (1982) source code, for example, is littered with sporadic running cycle count totals in addition to comments like 'WASTE 2 CYCLES' [51]. Other programmers adopted a more dynamic approach, making code changes until the picture rolled, then fixing the problem [45].

Needless to say, the limited memory along with the video timing constraints demanded that programmers write 2600 games in assembly. The 6507 was relatively slow, and there were only '76 machine cycles per [video] line' [60, page 4]. Given that 6507 instructions all take two or more cycles, there was no room for inefficiency.

### 2.4 Other Design Tradeoffs and Omissions

There were other conspicuous design tradeoffs and omissions that affected Atari 2600 programmers. In the above description, there is no mention of a BIOS or a ROM of any sort. The *user* of the 2600 completed the computer in a very literal way, in that the ROM for the machine was contained in the game cartridge the user plugged into the unit. There was no built-in set of ROM routines for a programmer to draw upon; even the reset vector data the 6507 first fetches upon power-up (i.e., the address the CPU should begin execution at when reset) is located externally to the 2600.

One important omission related to where a programmer happened to be. Atari was not in the habit of sharing technical documentation or software development kits with third-party developers. In fact, Atari went so far as to sue the first publishers of third-party Atari 2600 games [14, 27, 35]. For programmers who had worked for Atari making 2600 games, like the four who left Atari and founded Activision [14, 27], the lack of official technical information would not have been a problem. Other aspiring third-party developers had to start by reverse-engineering the Atari 2600 platform, however, like the company that produced Entombed [9].

While this whole-platform reverse engineering task seems daunting, and we do not intend to minimize it in any way, it is fair to consider how difficult it would have been for developers. Two-thirds of the 2600's innards were commodity chips; data sheets and other documentation for the RIOT and 6502 (if not the 6507 specifically) would have been readily available. The only real mystery was the custom TIA chip, but luckily the Rosetta Stone for understanding the TIA shipped with the Atari 2600. Code in game cartridge ROMs was not subject to any technical protection, and dumping and disassembling a ROM's contents would be neither difficult nor expensive. The 1977 game Combat, specifically, was a pack-in game for the Atari 2600 and was created to exercise the 2600's hardware as it was developed. As Montfort and Bogost put it, '*Combat* is practically a pure demonstration of the capabilities of the Atari VCS [2600], showing how they were intended to be used' [32, page 19]. Combat was not even a large ROM image to reverse engineer, weighing in at only 2 KiB. In short, reverse-engineering enough of the Atari 2600 to program it would appear to have been a tractable problem: the spartan nature of the platform lended itself to revealing its secrets.

That same simplicity did make it a challenge to program, though. It is intuitively easy to see why, given the limited memory, the slow processor, the real-time requirements, and the need to conceptualize and code each video frame in terms of individual lines. The programming challenges are noted by former 2600 programmers in interviews. John Harris said 'The way the machine had to be programmed was so obscure and so challenging that you had to write extremely optimal code' [19]; Doug Neubauer observed 'Using a 6502 to race an electron beam across the screen is a lot harder than having a full-screen bitmap to play with' [18]; Warren Robinett simply stated 'It was pretty challenging to do any game on it' [20]. Ultimately, the 2600's constraints shaped the design of the software that ran on it, like Entombed.

## 3 Case Study: Entombed

The physical artifact of Entombed (Figure 3) is a standard Atari 2600 cartridge, with a slightly more sculpted case than the usual Atari fare (e.g., Figure 4). But the interesting aspects lie inside, in the digital artifact.

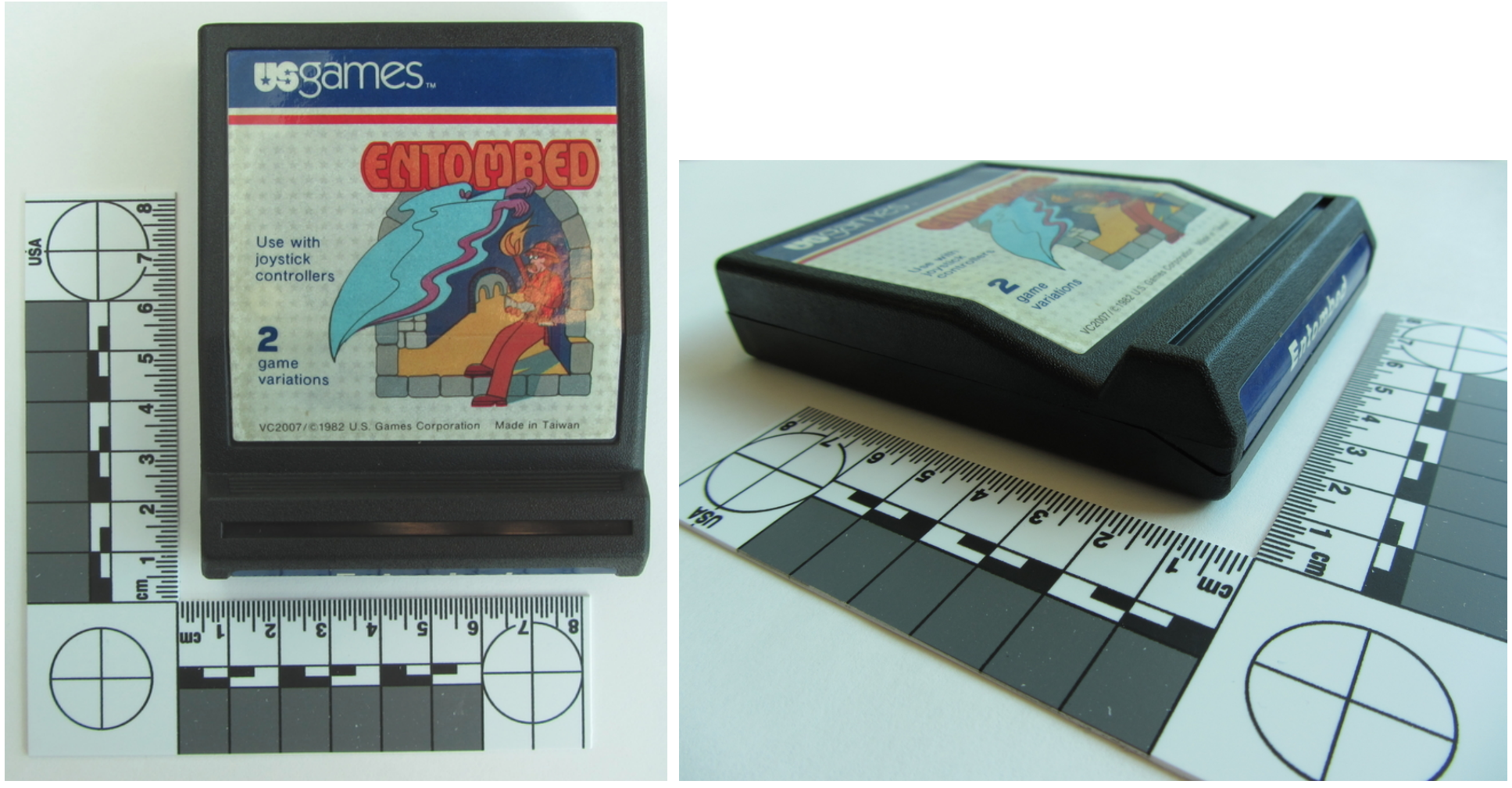

■ **Figure 3** Entombed, the physical artifact, front view (left) and side angle showing beveled case (right)

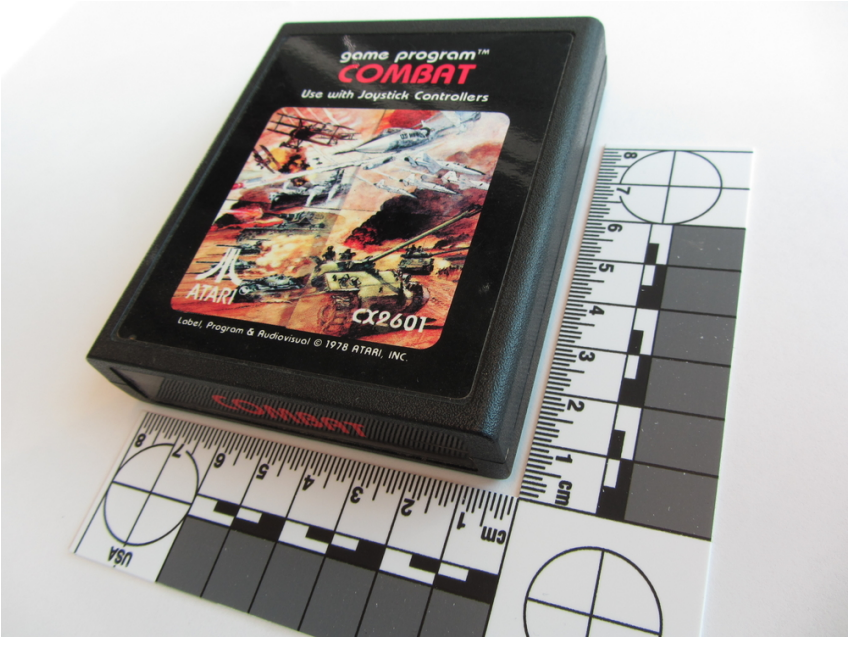

■ **Figure 4** Atari 2600 Combat cartridge for physical comparison

Before embarking on this case study of Entombed, it would be a reasonable question to ask why we chose this particular artifact over all other Atari 2600 games. The first author teaches a senior computer science course on retrogame implementation [8], and he was looking for a game that students could use for a reverse engineering assignment. The game had to have interesting features to study, yet at the same time be relatively unknown, because a more popular entry in the Atari 2600 canon would be likely to have a publicly available disassembly for students to find. The choice of Entombed was thus largely serendipitous. Given that archaeological finds can similarly rely on chance, this method of selecting Entombed seems apropos. Indeed, we would argue that the fact that there were so many interesting features within the 4 KiB ROM image of this *one* arbitrarily-selected game suggests that there may be many more things to find in other artifacts from the era.

Figure 5 shows what Entombed looked like. A randomly-generated maze gradually scrolled up the screen; if the "archaeologist" player character was forced to the top of the screen, or touched a zombie (not shown), the player would lose a life.

The maze shown in the figure already reveals some of the Atari 2600's architecture. The maze walls are represented by the TIA's 20 bits of 'playfield' [60] per line, which describe half the screen; the playfield can be either duplicated or reflected. The latter is done for Entombed, yielding the vertical symmetry of the maze. Another notable feature of the maze captured here is that it is unsolvable: there is no path by which the player character can move to the bottom of the screen. This would spell certain doom for the player, were it not for the free-floating rectangle in the upper right part of the screen. This is a "make-break," and acquiring these allows the player to make a new piece of maze wall or, more importantly, remove an existing piece of wall. Wall pieces added and removed take effect on both halves of the screen, again reflecting the symmetry. The fact that make-breaks are needed speaks to the algorithm used for maze generation, which we examine in Section 3.2. But first, we explain the methodology used in this work.

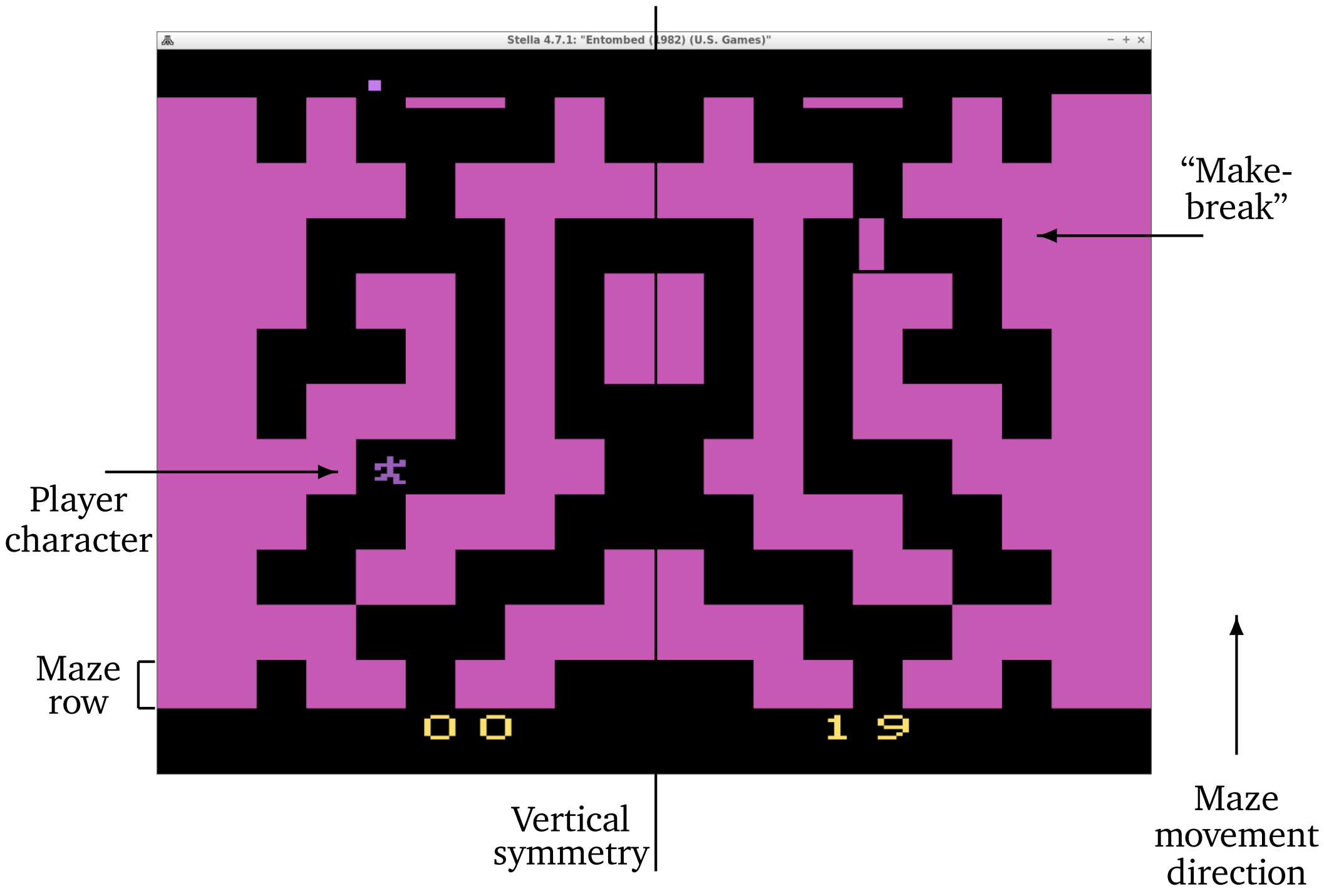

**Figure 5** Entombed game screen, highlighting design and gameplay aspects

### 3.1 Methodology

We began by manually reverse-engineering the relevant parts of Entombed's binary code, via both static and dynamic analysis using the Stella Atari 2600 emulator (version 3.9.3 on MacOS X for the initial analysis, and more recently Stella 4.7.1 on Linux along with MAME/MESS 0.191). Analysis of binary code without the involvement of the code's author has a long tradition in computer security (i.e., analyzing malicious code), up to and including analyzing its bugs; Spafford's analysis of the Internet Worm is a notable early example [53]. The correctness of our analysis is verified using reconstructions and objective comparison between our reconstructions and the real game, as detailed in Appendices B and C. We focused on the maze (and the pseudo-random number generator it relies upon) as the dominant, distinctive feature of Entombed, and additionally that permits comparison with those types of algorithms in games published contemporaneously [10].

In our setting, we can view reverse engineering as a type of archaeology, a means to discover if a digital artifact has interesting implementation features to study, and as critical background preparation for fact-checking or formal interviews with developers. It is not necessarily the case that developers retain source code that would avoid reverse engineering, either. In our experience, source code for early games is frequently lost or, if extant, may be legally encumbered or otherwise unobtainable, and for all these reasons we start with the binary code.

The mention of legal encumberment correctly suggests that there are copyright considerations when doing our research. Our work falls under research and education fair dealing exceptions (similar to fair use in the United States) of the Copyright Act of Canada [37]. However, this would not extend to distribution, meaning that it is not possible to share an archived binary image of Entombed or its full disassembly. Even linking to infringing material can run afoul of the law [48], and in any event there are multiple locations for game images. Instead, as games' binary images are quickly found with Internet search engines, we supply a means for ascertaining if a found image matches the one we used: a 4 KiB Entombed image with MD5 checksum 6b683be69f92958abe0e2a9945157ad5.

After reverse engineering, we turned to interviews to augment our understanding of Entombed's development. From this era, there is typically a very small set of people – frequently one person – responsible for game development, making the choice of interview subjects obvious. (We discuss the interview results and development practices further in Section 3.4.) However, our experience indicates that this small set of interview subjects can be hard to locate, unresponsive, or unfortunately deceased. We were lucky, in Entombed's case, to find two existing interviews, and we were able to contact Entombed's programmer for our own additional interview.[5]

Our oral history interview[6] was conducted as a semi-structured interview. We asked questions via email, making transcription accuracy not an issue, and the interview subject had the opportunity to review the final interview transcript prior to publication in our institutional repository [9]. Given the brevity and specificity of the interviews, it was trivial to identify the portions pertaining to the maze algorithm and the pseudo-random number generator, and consequently we have treated the interviews as source documents and incorporated all that material into Section 3.4.

There are cautions with respect to oral histories and accuracy [43], like a tendency to exaggerate one's contribution or importance. We argue that these cautions extend, in the case of programmers asked about long-past code, to technical details. Simply too much time has elapsed to recall reliable technical information, and we see evidence of vague and conflicting memories in Section 3.4. All is not lost, though, because many hidden truths that a programmer left behind reside in the code.

[5] Interview conducted with ethics approval from the University of Calgary Conjoint Faculties Research Ethics Board, File REB16-1235.

[6] See [43] for a good general oral history introduction.

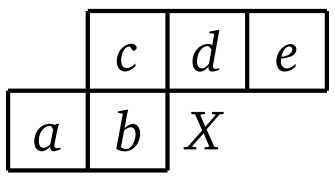

**Figure 6** Maze wall context used to select a new bit

### 3.2 Maze

As mentioned, we reverse engineered the somewhat convoluted assembly code for the maze generation algorithm used by Entombed. The maze is dynamically generated row by row:[7] as the maze scrolls up, bits are produced representing the newly revealed row at the bottom of the screen. Due to vertical symmetry, at most the 20 bits of the playfield need be generated, but in fact it is much less than that.

The 20 bits of the TIA's playfield are split across three TIA registers, one of which holds the leftmost four bits of the playfield [60]. Entombed's maze always has a wall on the left-hand (and through reflection, right-hand) side that uses exactly four bits. The maze generation thus is left to produce 16 bits. However, each playfield bit of the maze is duplicated in Entombed, meaning the generation algorithm only has to select eight bits to represent a new maze row.

The way each bit is chosen is shown in Figure 6. For each bit $X$ that represents a piece of maze wall, from left to right, five bits of wall context are extracted, the two immediately to the left of $X$ ($a$ and $b$), and the three above $X$ ($c$, $d$, and $e$). Initially, $a \ldots c$ would be located in the omnipresent left-hand wall, and the algorithm initializes them to $a = 1$, $b = 0$, and $c$ to a bit drawn from the pseudo-random number generator we examine in the next section. For the eighth bit generated, $e$ would be across the line of vertical symmetry and provide the same information as $d$, and $e$ is instead set to another pseudo-random bit for this final step.

The remaining question is exactly how $a \ldots e$ are used to choose $X$, and there is no completely satisfactory answer. The five wall context bits index into a mysterious 32-byte lookup table, from which is drawn one of three results for $X$: wall, no wall, or a pseudo-random decision. The mapping is shown in Table 1, and no obvious patterns in the mapping are apparent. To increase our confidence, we represented the table as a Karnaugh map (choosing a number of different representations for the trinary value of $X$), but the algebraic solutions did not reveal any simple patterns of note. Our conclusion is that the table values were manually chosen, or manually tuned, by the maze algorithm designer.

As an example, Figure 7 steps through the algorithm, illustrating how the last two maze rows in Figure 5 may be produced. Notice how the maze wall context "window" shown in Figure 6 slides from left to right one bit at a time as each maze row is generated.

Following each maze row's generation, Entombed performs what we refer to as postprocessing: looking for one of two patterns in recently-produced maze rows and, finding either pattern, making an adjustment to the last-generated row. The first pattern is encountered extremely rarely in our testing,[8] where the last 11 maze rows all have a 0 in the leftmost column. The second pattern is more frequent but more difficult to describe, where the last 7 bits in the rightmost column (i.e., the center of the maze) all match the *ninth* last rightmost bit. Because the eighth bit is skipped, the second pattern actually catches four different combinations of the last 9 rightmost bits. In any case, finding either pattern causes all or part of the last-generated maze row to be reset to 0, effectively breaking out of the detected pattern.

[7] To avoid ambiguity, we refer to maze *rows* to distinguish them from screen lines. Each maze row is represented by multiple lines onscreen (see Figure 5).

**Table 1** Mystery table mapping used for maze generation

| $a$ | $b$ | $c$ | $d$ | $e$ | $X$ |
|---|---|---|---|---|---|
| 0 | 0 | 0 | 0 | 0 | 1 |
| 0 | 0 | 0 | 0 | 1 | 1 |
| 0 | 0 | 0 | 1 | 0 | 1 |
| 0 | 0 | 0 | 1 | 1 | random |
| 0 | 0 | 1 | 0 | 0 | 0 |
| 0 | 0 | 1 | 0 | 1 | 0 |
| 0 | 0 | 1 | 1 | 0 | random |
| 0 | 0 | 1 | 1 | 1 | random |
| 0 | 1 | 0 | 0 | 0 | 1 |
| 0 | 1 | 0 | 0 | 1 | 1 |
| 0 | 1 | 0 | 1 | 0 | 1 |
| 0 | 1 | 0 | 1 | 1 | 1 |
| 0 | 1 | 1 | 0 | 0 | random |
| 0 | 1 | 1 | 0 | 1 | 0 |
| 0 | 1 | 1 | 1 | 0 | 0 |
| 0 | 1 | 1 | 1 | 1 | 0 |
| 1 | 0 | 0 | 0 | 0 | 1 |
| 1 | 0 | 0 | 0 | 1 | 1 |
| 1 | 0 | 0 | 1 | 0 | 1 |
| 1 | 0 | 0 | 1 | 1 | random |
| 1 | 0 | 1 | 0 | 0 | 0 |
| 1 | 0 | 1 | 0 | 1 | 0 |
| 1 | 0 | 1 | 1 | 0 | 0 |
| 1 | 0 | 1 | 1 | 1 | 0 |
| 1 | 1 | 0 | 0 | 0 | random |
| 1 | 1 | 0 | 0 | 1 | 0 |
| 1 | 1 | 0 | 1 | 0 | 1 |
| 1 | 1 | 0 | 1 | 1 | random |
| 1 | 1 | 1 | 0 | 0 | random |
| 1 | 1 | 1 | 0 | 1 | 0 |
| 1 | 1 | 1 | 1 | 0 | 0 |
| 1 | 1 | 1 | 1 | 1 | 0 |

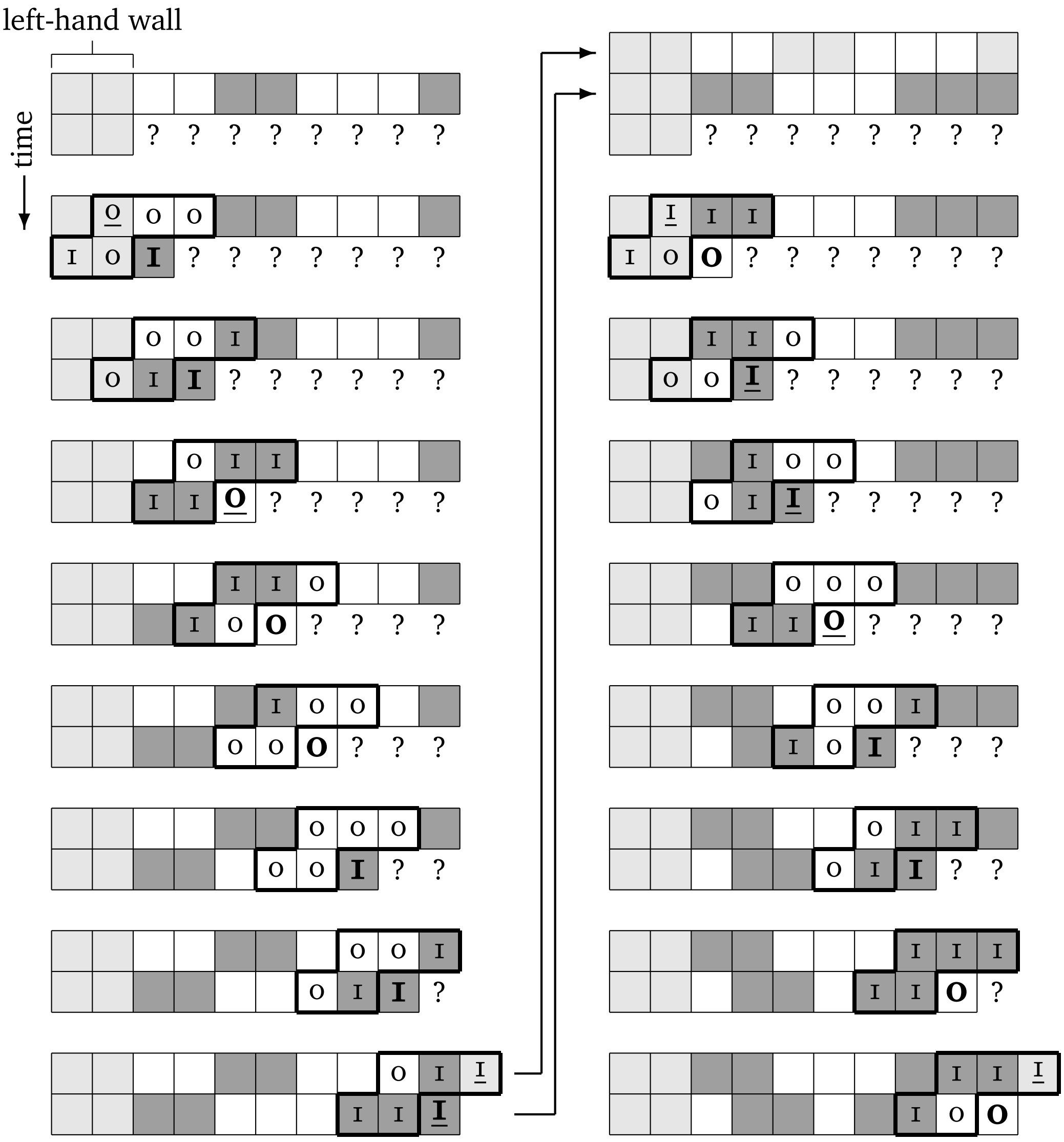

**Figure 7** Maze row generation example (generated bits are bold, PRNG bits are underlined, and lighter gray boxes are 1-bits present but unused)

*Ad hoc* postprocessing aside, there were a number of maze-based games in the early days of computer games, and it is instructive to undertake some comparative archaeology. Here we consider the maze algorithms used in Amazing Maze (1978), 3D Labyrinth (1982), and Rogue (1993)[9] [10]. Including Entombed's maze algorithm in the mix, the common theme is a lack of commonality: there is a diversity of algorithms even amongst those that have solvable mazes. It would be easy to dismiss this negative result out of hand, but the same thing is seen even in different areas of retrogame implementation, areas as seemingly mundane as text and line representations [10]. There seems to be a great deal of programmer – of human – creativity buried in the code of retrogames, underscoring our earlier point that these are a rich set of artifacts to examine.

Finally, we built a reconstruction of the maze-generation algorithm in Python to test our reverse-engineered result. A sample of its output is included in Appendix A, and the Python code is in Appendix B along with details of how we verified its correctness. Normally an experimental reconstruction would be compared with the original to ensure perfect fidelity [11], but in Entombed this is somewhat challenging. The pseudo-random number generator that the maze algorithm uses is shared with other elements of the game that need random numbers, and the random number generator itself has some quirks.

### 3.3 Pseudo-Random Number Generator and Code Reuse

The variety of patterns shown in the maze generation relies on the pseudo-random number generator (PRNG) in the game code. Its 6507 disassembly is shown in Listing 1, with references to the consecutive zero page RAM locations `$dd`, `$de`, `$df`, and `$e0` replaced with `W`, `X`, `Y`, and `Z` respectively. As the code is important to our results, and not all readers are likely to be familiar with 6507 assembly, we will explain it in some detail. A reconstruction in Python is provided in Appendix C along with verification details.

A good way to understand this code is by thinking of `W` through `Z` as representing two big-endian 16-bit values, where `WX` acts as the last number generated by the PRNG

[8] In the verification data described in Appendix B, we saw the first pattern only 10 times in 300,000 maze rows; the second pattern appeared 4324 times in 5000 mazes.
[9] This is a later release of Rogue; early versions did not have the mazes.

**Listing 1** Entombed's PRNG implementation

```
1  LBCA5:
2          LDA     W
3          STA     Y
4          LDA     X
5          STA     Z
6          ASL
7          ROL     W
8          ASL
9          ROL     W
10         CLC
11         ADC     Z
12         STA     X
13         LDA     #$00
14         ADC     W
15         CLC
16         ADC     Y
17         STA     W
18         LDA     #$00
19         INC     X
20         ADC     W
21         STA     W
22         RTS
```

(and initially, the PRNG seed), and `YZ` is a temporary value. The 6507 is only able to operate on 8-bit values, and the 16-bit operations must be decomposed accordingly. Additionally, the 6507 has a single accumulator, `A`, that is used extensively here; some instructions name it explicitly (`LDA`, `STA`), others implicitly (`ASL`, `ADC`).

The load and store instructions in Lines 2–5 begin the PRNG by copying the value of `WX` into `YZ`, leaving the least-significant byte (LSB) of `WX` in the accumulator. Lines 6–7 multiply `WX` by two, by left-shifting its in-accumulator LSB left (into the carry bit), then rotating `W` left: the 6507's rotate is a 9-bit rotation through the carry bit. The multiplication by two is then repeated in Lines 8–9.

Lines 10–12 start the 16-bit addition `WX+YZ` by adding the LSB of `WX`, which is still in the accumulator, to the original LSB value as copied to `Z`. The 6507's only add instruction is an add with carry, making management of the carry bit (here, `CLC` to clear the carry) important. The resulting value is finally stored out of the accumulator back into `X`.

Half of the 16-bit addition is complete at this point. The most-significant byte (MSB) of `WX` is computed by Lines 13–17 in two parts. First, 0 is added to the MSB in `W`, propagating any carry from the LSB. Second, `Y` is added to the possibly-updated `W` value in the accumulator, and then stored into `W`.

The last step before returning, in Lines 18–21, is a 16-bit increment of `WX`, incrementing the LSB `X` in memory and propagating the carry through to `W` by adding 0 (and the carry). The overall calculation is

$$\mathsf{WX} = 4 \times \mathsf{WX} + \mathsf{WX} + 1$$

which, combining terms and taking the 16-bit nature of the computation into account, is actually

$$\mathsf{WX}_{n+1} = (5 \times \mathsf{WX}_n + 1) \mod 65536,$$

the form of a linear congruential PRNG algorithm [28] with constants 5, 1, and 65536. The implied modulo from the 16-bit wordsize is suboptimal in terms of the PRNG's generated numbers [28], but is not uncommon to find in retrogame code [10].

The reason why it was necessary to go into depth on the PRNG code was because of what happens at the very end: it has a bug. Lines 18–21 should increment `WX` by one, and they would *if* the `INC`rement instruction modified the carry bit. If that were the case, the LSB `X` would be incremented, and the carry would be correctly propagated to the MSB `W` with the add of 0. Instead, the carry is incorrectly propagated with a value from earlier in the PRNG. According to our tests, the resulting PRNG is hampered in that a non-buggy version of this PRNG would have a full period from all starting seed values, whereas the maximum number of distinct values we saw with the buggy version from any seed was 1200.[10] Still, in a game environment where a "good enough" PRNG is sufficient, this is not a show-stopping problem. An Entombed player would definitely not notice any effects from the bug, as a plausible pseudo-random sequence is still generated. Would a programmer notice the bug? We suspect not: the PRNG works, and we argue that a programmer would mentally abstract away that routine once written. Our tests showed that 50.3% of generated values corresponded to the non-buggy values, and in each case where they failed to coincide, the high byte (`W`) is only off by one.

What this bug does give us, however, is a very distinctive signature for this PRNG code. From an archaeological standpoint, we can use this to identify programmers' code reuse activity. We downloaded a corpus of 533 Atari 2600 game ROM images from AtariAge [6] and searched for the PRNG code in all the ROMs. The exact zero page addresses of `W`, `X`, `Y`, and `Z` were treated as wildcards in case the variables moved around in different games.

Besides Entombed, we found this code in five other games. Three (M.A.D., Raft Rider, Towering Inferno) were also published by US Games; another, Q*bert, has people credited that intersect with the US Games games [7]; the last one, Angriff der Luftflotten, has almost no available information to work with, but appears to be a minor variant of M.A.D. Regardless, the common buggy PRNG gives very strong evidence of code reuse. It seems clear that the PRNG code either originated elsewhere and was copied into Entombed, or was copied from Entombed to other games. Who might have done this?

### 3.4 Design and Development

Early console game programmers were frequently uncredited for their work [14, 27]. Even a cursory examination of early games and their packaging is sufficient to verify the lack of programmer credit, and it was the motivating factor behind Warren Robinett's famous Easter egg in Adventure (1980) for the Atari 2600 [44, 41:19]. So instead, we begin the search for authorship in non-primary sources. Wikipedia, circa 2016, claimed that Entombed was 'written by Tom Sloper and Jeff Corsiglia' [58], information also found on the game's MobyGames entry [31]. Most recently, the Wikipedia entry has been altered to read 'designed by Tom Sloper and programmed by Paul Allen Newell' [59]. As it turns out, all are wrong in whole or part.

[10] We chose this as a more meaningful metric rather than the period because the buggy PRNG would rarely return to the seed value.

**Table 2** Entombed development credits

| Source | Role | Credited |
|---|---|---|
| Newell interview [54] | Maze algorithm | Duncan Muirhead, Paul Allen Newell |
| | Programming | Steve Sidley |
| | Game design | Tom Sloper |
| Sloper interview [39] | Programming | Steve Sidley |
| | Game design | Tom Sloper |
| | Original design | Jeff Corsiglia |
| Sidley interview [9] | Programming | Steve Sidley |
| | Original design | Jeff Corsiglia, Tom Sloper |
| | Maze algorithm | not explicitly named |

Still, as a starting point this proved to be helpful. We found the transcript of an interview with Tom Sloper [39], and armed with information from there found another interview with Paul Allen Newell [54]. Both sources agreed that the programmer of Entombed was a Steve Sidley, whom we were able to locate and interview [9]. Combining all these sources together reveals the authorship information in Table 2. The maze algorithm came first (Muirhead and Newell), a proof of concept of the maze generator was programmed for the 2600 (Newell), a game was designed from it (Corsiglia, then Sloper), and the game was written (Sidley).

The early days of game programming were the age of the so-called "bedroom programmer," a solo developer; this is reflected in numerous oral histories and what credits are known for early games. This was the case even inside relatively large companies, as Warren Robinett recollected from his time at Atari [20]: 'Each 2600 game was designed entirely by one person.' It is therefore notable that at a time when a usual scenario would be one programmer, one game, that *five* different people were involved in one way or another with the development of Entombed.

We also learn more about the development and evolution of the maze algorithm. As Newell tells it [54], 'Duncan and I went out for a beer and ended up coming up with this "problem" of wondering whether one could generate an endless maze that always had a solution' and that 'We worked out the algorithm and [...] I spent a weekend coding something up.' The assertion that the generated maze was solvable is at odds with the final maze in Entombed, of course, as Figure 5 attested to. The original maze algorithm apparently had further properties: for instance, Newell went on to say that the algorithm was parameterized to give an adjustable level of difficulty.[11]

Sidley's recollection of the algorithm is slightly more colorful [9]:

> 'The basic maze generating routine had been partially written by a stoner who had left. I contacted him to try and understand what the maze generating algorithm did. He told me it came upon him when he was drunk and whacked out of his brain, he coded it up in assembly overnight before he passed out, but now could not for the life of him remember how the algorithm worked.'

Sidley also observed that the maze code was uncommented, and when asked about the 32-byte table said 'It was a mystery to me too, I couldn't unscramble it. I just used it to generate the new row at the bottom of the screen.'

Regardless of which version of events is followed, it seems fair to say that some level of intoxication was involved in the development of the maze algorithm.[12] Newell's claim that the original maze was solvable is certainly possible: Eller's algorithm, for instance, generates solvable mazes a row at a time [12]. (And, coincidentally, was invented in 1982 [12], the year Entombed was released.)

Sidley's interview [9] also gives glimpses of the development environment and practices. Their need to reverse engineer the Atari 2600 in lieu of documentation was already mentioned, but it is striking that Sidley – despite having completed a Master's degree in computer science from UCLA just prior – had never programmed in assembly language before. This is corroborated by Newell, who mentioned giving Sidley a 6502 primer [54]. Sidley did not explicitly remember any code reuse between games, although recalled there were some PRNGs around the company. He did seem to harbor some nostalgia for that time [9]: 'Those of us who programmed 6502 for those games have a special bond. Like climbing Everest without oxygen.'

It is worth mentioning that we did not ask Sidley about the PRNG bug. In our view, three things weighed against it: the difficulty of spotting the bug; the ease with which the bug could have been fixed had it been known; the unreliability of memory regarding long-past technical minutia.

## 4 In Context: Archaeology and Archaeogaming

To understand how programming and games fit into the field of archaeology, and to separate it from inaccurate Hollywood depictions, we first need to step back and examine what archaeology *is*.

Broadly speaking, archaeology can be defined as 'the study of the ancient and recent human past through material remains' [49]. The material remains used within archaeology can include physical artifacts, features, architecture, and landscapes as well as biological and ecological materials [42]. Archaeology is not limited to the physical: there is also significant scope for blending additional disciplines such as heritage or history to explore intangible elements, textual material, experimental reconstruction, and consideration of ephemeral entities such as experience and emotion.

[11] We have made several attempts to contact Newell for more information, to no avail.

[12] Perhaps this is the reason why the code seemed convoluted, as we mentioned in Section 3.2.

Archaeology's focus on material remains separates it from closely related fields such as history, in which the analysis of textual documents and records are the primary methods used to unpick the past. The use of material remains allows archaeologists to explore data that either could not be captured via text or otherwise was suppressed, omitted, or not deemed necessary to be captured in written forms. In other words, '[a]rchaeology shows us things we are not normally meant to see' [36, page 39].

At its core, archaeology is concerned with what things are, when they are from, what they functioned as, how they were made, the impact that they had on both individual humans and wider societies, as well as the impact that humans had on these things. In short, it is a discipline that is concerned with identifying and understanding the intertwined relationship between 'things' and the human ways of life that include them [23]. Here we have taken a blended approach, delving deeply into the archaeology of Entombed through material analysis and the excavation of code, using interviews to contextualize and explore these findings in greater detail. This section places our work in an archaeological framework, and makes a case for the value of this approach for exploring and understanding computer games.

Computer games, such as Entombed, are an example of material culture from the recent human past – physical and digital artifacts born from human interactions with code, art, audio, and industrial design which go on to have additional lives in the archaeological record through play and discard [13]. As a type of material culture, it stands to reason that computer games can be examined and understood through archaeological lenses. Indeed, several computer game artifacts have been found in the physical archaeological record, such as the E.T. cartridges excavated in Alamogordo, New Mexico [41]. In cases of physical excavation, the traditional archaeological toolkit of trowels and shovels has proven effective. However, when faced with the question of how to excavate and understand the construction and operation of computer games under the hood, the traditional archaeological toolkit faces some significant challenges; *archaeogaming* is an emerging branch of the archaeological discipline which is dedicated to exploring such challenges.

Archaeogaming has been defined as 'the archaeology both in and of digital games' [40, page 2], a field of study that ranges from exploring representations of archaeology within a game to excavating code, and even creating games to explore archaeological methods and theories. At the core of archaeogaming is the idea that computer games 'provide landscapes and objects that are productive for archaeological investigations of digital materiality' [33, page 9], a vastly important area of study given the commercial and social impact that computer games have had on recent human past and present.

In analyzing the visual and narrative aspects of Entombed via archaeogaming, we find tenuous links to archaeology: the player character is described as being an archaeologist attempting to traverse a zombie-infested maze. The second author's experience as an archaeologist, as well as extended literature on the topic of archaeology, would indicate that this representation is inaccurate, though potentially has a popular culture basis in the adventure-filled fictional literature spawned by the grand stories of the antiquitarian era. However, as we shift our archaeogaming approach away from the rendered aspects, to what is happening under the hood of Entombed, the archaeogaming approach revealed information beyond what could be observed in textual documents or surface readings of the game through play.

Material culture, including computer games, is crafted through human interaction with a medium. Through this interaction, the constraints of the medium have a hand in shaping the potential space of what can be made. In traditional archaeology, for example, the composition and properties of different rocks afforded the flintknapper[13] certain opportunities and precluded the production of other types of stone tools. Through archaeological observation and engagement, we can come to understand the role that physical mediums have on shaping production and human interaction. In analyzing computer games archaeologically, we put aside microscopes and chemical composition testing in favor of design schematics and engineering approaches. As presented in Section 2, the Atari 2600 had a number of physical, intrinsic constraints which shaped the potential space of Entombed through the affordances of how code could be constructed and executed.

In addition to exploration of the medium, an archaeological approach is interested in exploring how something was made, with what, and by whom. The marks of human activity on a medium can be found, for example, in the tool marks or impressions a potter leaves in their wares. These physical tool marks and impressions are a kind of signature that can be analyzed through scientific method, comparison to existing specimens, and archaeological interpretation. The analysis of these signatures can tell us a lot about the tools used in construction, the person or peoples involved in creating them, as well as the conditions and broader society that frame their construction. Given that programming is a human activity, it stands to reason that by excavating and analyzing code we can find similar marks that provide clues not only as to how it was made but also who made it.

Often in archaeology we find surprises, differences between what we expect to find and the information that we subsequently uncover and interpret [26]. In the case of Entombed, analysis of the code revealed abnormalities that had been replicated across other games, a kind of "maker's mark" that went on to have wider cultural significance through its replication. Such replication would traditionally represent a single maker (or group of makers working under the same approach) implementing the same process. While this may still be the case with Entombed, the programmer's medium allows for instant and perfect duplication, meaning that the evidence of the maker's mark can be transferred identically across iterations. This represents an interesting set of challenges for archaeology. Traditional archaeological methods such as seriation and dating can help unravel the temporal sequence of these marks; however, further investigation into their meaning and context requires the augmentation of archaeology through computer science, anthropology, or history. Given the age of computer games, we are in the fortunate position of being able to interview and interact with the creators, a situation that many Roman or paleolithic archaeologists wish they were in.

[13] A flintknapper is the technical term for a person who creates stone tools.

In this paper, the fragments of human activity, manifested in code, posed questions of authorship and process that otherwise would have been hidden from sight. Then, comparison across other Atari 2600 games showed a wider societal framework of reuse, and through augmenting these archaeological approaches with interviews we were able to generate layers of meaning around these entities. These approaches provided valuable – though at times conflicting – information, a reminder that human memory can be fallible, written history incomplete, and that material remains can only ever capture part of the story.

Computer games are a type of material culture [40], emergent from the human activities of programming and design. Using archaeological approaches allows for the data and stories that result from the medium and production process to be explored meaningfully, allowing unseen elements to be found, explored and given meaning. However, the position of computer games in straddling the physical (hardware) and the digital (software) poses some interesting challenges to the archaeological discipline, and necessitates the development of new toolkits and approaches to facilitate effective archaeological exploration. As demonstrated here, archaeology provides an effective framework for exploring the complex entanglement between computer games as material culture and the human activities involved in creation of, interaction with, and disposal of things; however, for archaeological approaches to be meaningful in this context, collaboration with computer science approaches is required.

## 5 Related Work

"Archaeology" is a term that is compelling, thanks to pop-culture depictions of archaeologists, and also one that is easy to co-opt or at least use metaphorically. In terms of software, a typical definition is that software archaeology is the understanding of legacy code [24, 46], and consequently the archaeology label rears its head occasionally in software engineering research. For instance, the 2010 ACM/IEEE ICSE conference had three papers presented in a 'software archaeology' session [25]. What is unclear about the work done under this moniker is how often an actual archaeologist is involved; archaeology is, of course, a discipline much older than computer science, and we argue that it is a natural opportunity for interdisciplinary work as we have begun here.

Further, in computer science, a general solution is often seen as better or more elegant than a specific one. While we are aware that the arguments we make and the techniques we use here could be construed as a case for software archaeology in general, we make the somewhat counterintuitive claim that a specific focus on *game* archaeology is advantageous. To explain why, we need to step back and examine the landscape of allied areas of study.

As artifacts with cultural relevance, games have been subject to organized academic study in the humanities for almost two decades; Aarseth claims 2001 as 'Year One' of the field of game studies [1]. They are not alone: there is active work in game history (e.g., [30]), and an entire area called "platform studies" dedicated to 'the investigation of underlying computing systems and how they enable, constrain, shape, and support the creative work that is done on them' [32, page vii] (which, despite the general description, over half the books on the topic at present are game platforms). In other words, there is already considerable ongoing research and expertise to draw upon by virtue of staying within the realm of computer games, not to mention narrowing the otherwise dizzying range of artifacts that might be chosen to study.

Where existing work on games can fall short is in technical depth and rigor, and even a deliberate exclusion of technical methods. For example, one book touting itself as 'a comprehensive introduction to the field of game studies' [15] does not even mention studying game code as a type of analysis [15, page 11]. (Others are not quite as exclusionary: Fernández-Vara notes that a game's 'technological context' may be of import, with reference to platform studies [17].) One might imagine that computer history, another possible allied field, would be immune to this problem, and indeed there are attempts to capture technical history, such as the intermittent History of Programming Languages conferences [2]. But, as Haigh notes, 'the technical history of computer science is greatly understudied' [21, page 43].

An all-technical approach to studying games with a heavy computer science bias is certainly possible. Retrogame archaeology was previously defined as 'understand[ing] the tools, techniques, and technology used in old games' implementation [...] and placing them in a broader, modern technical context' [10, page 206]. However, this myopic technical view omits the human element which is needed to fully understand the context of development, and is something added through an interdisciplinary approach involving archaeologists.

Specific to maze generation algorithms, they fall under the broader umbrella of procedural content generation (PCG). While there are a number of treatments of PCG [22, 50, 52, 56], they tend to focus on recent developments rather than past techniques. The only work we are aware of examining historical PCG in technical detail is our own [10], and Entombed's unusual (and possibly unique) algorithm adds to that body of knowledge.

## 6 Conclusion

Computer games are a not just a technical product; they are a form of material culture which can be examined through archaeological lenses. Here, we swapped the traditional archaeological tools of shovels and trowels for digital counterparts, drilling down into the material remains of Entombed's code to expose artifacts of the human process of programming. Comparisons to other maze-based games revealed a diversity of approaches to maze generation, a finding which highlights the uniqueness of these artifacts as well as the human creativity involved in the design and programming of computer games. Additional comparisons to a corpus of Atari 2600 games revealed a practice of code reuse, a finding which touched upon the wider cultural landscape of computer game creation during this era.

As artifacts of the recent past, computer games also offer archaeologists the chance to augment, compare, and contrast material culture with interviews and history. Here, this practice provided significant insights as well as affirming a constant struggle between anthropology, archaeology, and history: that humans, material remains, and written texts can all tell very different stories.

Finally, the position of computer games in straddling the physical (hardware) and the digital (software) poses a set of domain-specific challenges to archaeology that archaeogaming seeks to untangle and, as this paper demonstrates, for which collaboration with computer science may prove particularly fruitful.

The limitation of this work is that the methodology across computer science and archaeology is not yet as fully and deeply integrated as it will become in future: this is only the start of an interdisciplinary discussion. Through our case study, we are beginning to see how the technical study of programming artifacts can be enhanced though archaeological, human context and interpretation; in the other direction, the technical ability to understand and reverse engineer code is of benefit to archaeologists coming to terms with an increasing glut of contemporary digital artifacts. The interdisciplinary archaeology-computer science combination sets the stage for novel approaches to study programming and code, in ways that are contextual to the people who produced it, and appropriate for challenges posed by aging software.

**Acknowledgements** The first author's work is supported in part by a grant from the Natural Sciences and Engineering Research Council of Canada. Thanks to the archive staff at the Strong National Museum of Play for access to References [16] and [51].

## A Maze Generator Reconstruction Output Sample

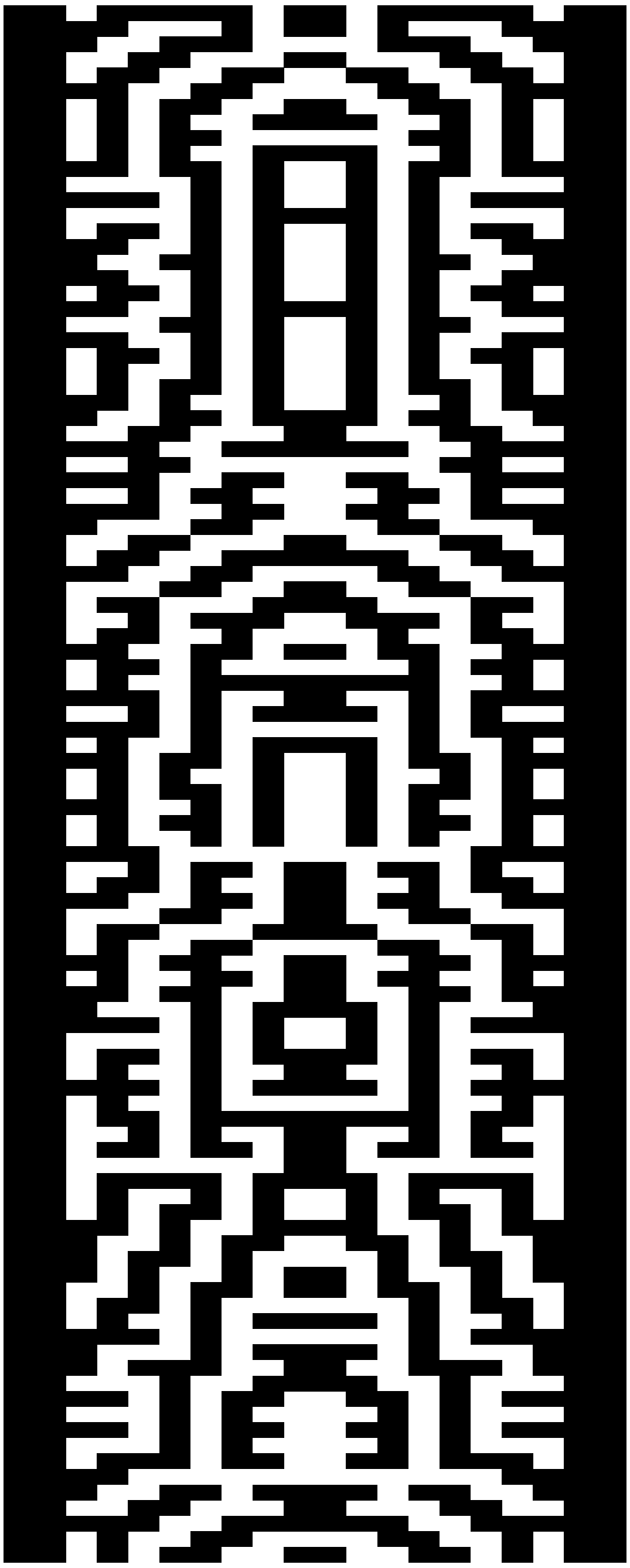

## B Maze Algorithm Reconstruction and Verification

Listing 2 shows our Python reconstruction of Entombed's maze-generation algorithm.

**Listing 2** Python reconstruction of Entombed's maze-generation algorithm

```
import random

def getrandombit():      return random.randint(0, 1)
def leftrandombit():     return getrandombit()
def rightrandombit():    return getrandombit()
def midrandombit():      return getrandombit()
def generated(x):        pass

def prrow(seed):
        PF12 = ''                       # "PF" == 2600's playfield
        for i in range(8):
                if seed & 1:
                        PF12 = 'XX' + PF12
                else:
                        PF12 = '__' + PF12
                seed >>= 1
        PF012 = 'XXXX' + PF12
        print PF012, PF012[::-1]

# the mystery table from Entombed

MAGIC = {
        (0b00, 0b000):  1,
        (0b00, 0b001):  1,
        (0b00, 0b010):  1,
        (0b00, 0b011):  None,           # None == random bit
        (0b00, 0b100):  0,
        (0b00, 0b101):  0,
        (0b00, 0b110):  None,
        (0b00, 0b111):  None,

        (0b01, 0b000):  1,
        (0b01, 0b001):  1,
        (0b01, 0b010):  1,
        (0b01, 0b011):  1,
        (0b01, 0b100):  None,
        (0b01, 0b101):  0,
        (0b01, 0b110):  0,
        (0b01, 0b111):  0,

        (0b10, 0b000):  1,
        (0b10, 0b001):  1,
        (0b10, 0b010):  1,
        (0b10, 0b011):  None,
        (0b10, 0b100):  0,
        (0b10, 0b101):  0,
        (0b10, 0b110):  0,
```

```
    (0b10, 0b111):  0,

    (0b11, 0b000):  None,
    (0b11, 0b001):  0,
    (0b11, 0b010):  1,
    (0b11, 0b011):  None,
    (0b11, 0b100):  None,
    (0b11, 0b101):  0,
    (0b11, 0b110):  0,
    (0b11, 0b111):  0,
}

def rowgen(lastrows):
    # prepend and append random bits to last row
    lastrowpadded = leftrandombit()
    lastrowpadded <<= 8
    lastrowpadded |= lastrows[-1]
    lastrowpadded <<= 1
    lastrowpadded |= rightrandombit()

    # last two bits generated in current row, initial value = 10
    lasttwo = 0b10

    newrow = 0

    # iterate from 7...0, inclusive
    for i in range(7, -1, -1):
        threeabove = (lastrowpadded >> i) & 0b111

        newbit = MAGIC[lasttwo, threeabove]
        if newbit is None:
            newbit = midrandombit()
        newrow = (newrow << 1) | newbit

        lasttwo = ( (lasttwo << 1) | newbit ) & 0b11

    # hook for verification
    generated(newrow)

    # now do postprocessing
    lastrows.append(newrow)
    lastrows = lastrows[-11:]

    # postprocessing condition 1
    history = [ b & 0xf0 for b in lastrows ]
    if 0 not in history:
        if sum( [ b & 0x80 for b in history ] ) == 0:
            lastrows[-1] = 0

    # postprocessing condition 2
    history = [ b & 0xf for b in lastrows[-7:] ]
    if 0 not in history:
        comparator = 0
```

```
        if len(lastrows) >= 9:
            comparator = lastrows[-9]
        if sum( [ b&1 for b in history ] ) == (comparator&1)*7:
            lastrows[-1] &= 0xf0

    prrow(lastrows[-1])
    return lastrows

def mazegen():
    lastrows = [ 0 ]
    while True:
        lastrows = rowgen(lastrows)

if __name__ == '__main__':
    mazegen()
```

For the reasons explained in Section 3.2, it is difficult to verify the reconstruction's accuracy due to PRNG values being used (and therefore perturbed) elsewhere in the game. However, we were able to verify Listing 2 in the following manner.

We began by creating a patched version of the Entombed binary, with three changes applied, as shown in the table below. These patches keep the player moving continuously downwards without colliding with maze walls or zombies, permitting us to gather maze data from Entombed for an unending number of mazes.

| Address | Original Value | New Value | Reason |
|---|---|---|---|
| $b0d5 | $20 $18 $ba | $ea $ea $ea | Disable player-zombie collisions |
| $b59a | $01 | $fe | Disable player-playfield collisions |
| $b6a4 | $b0 | $90 | Force player movement downward |

We then ran the patched image in MAME (MESS) 0.191 as follows; maze.script is a debugger script (Listing 3) that instruments the patched binary to capture in-game maze generation data.

```
mess64 -debug -debugscript maze.script -log -window a2600 -cart Entombed
    ↪ -patched.bin
```

The instrumentation captures, in particular, the PRNG bit values that Entombed uses while generating the maze, as well as the generated result for each maze row. If our algorithm reconstruction, when fed those same PRNG bits, produces the same maze rows (without any PRNG bits left over), then our reconstruction matches the original. We compare the generated maze rows prior to the maze algorithm's postprocessing, because Entombed's postprocessing step actually works directly on the playfield register data, making direct comparison between postprocessed values not possible. However, we do this without loss of generality, as any maze modifications due to postprocessing affect maze row values in subsequent maze rows, which we do verify.

**Listing 3** Maze instrumentation script

```
bpset b0a1,1,{ logerror "game_start␣%02x%02x\n", b@dd, b@de; g }
bpset b5b9,1,{ logerror "maze_end\n"; g }
```

```
bpset b427,1,{ logerror "new_maze_row\n"; g }
bpset b494,1,{ logerror "postprocessing␣1\n"; g }
bpset b4ad,1,{ logerror "postprocessing␣2\n"; g }
bpset b464,1,{ logerror "maze_row␣%02x\n", b@87; g }
bpset b447,1,{ logerror "prng_bit_left␣%x\n", P&1; g }
bpset b461,1,{ logerror "prng_bit_right␣%x\n", P&1; g }
bpset b4ee,1,{ logerror "prng_bit_mid␣%x\n", P&1; g }
g
```

We captured five logs using MAME, each with 1000 mazes generated (300,000 maze rows in total, since each maze has 60 rows). Using a test harness for automated comparison, we have verified that our algorithm reconstruction matches the game output exactly. We have additionally verified that both postprocessing cases were seen in the logs, and that all entries in the “mystery table” were used.

## C PRNG Reconstruction and Verification

Listing 4 shows Python code that reconstructs the output from Entombed’s PRNG. Its overall structure can be understood in terms of the linear congruential generator equation given earlier: the multiplication by 5 (Line 9), and the addition of 1 (Line 16). However, the bug in Entombed’s PRNG complicates matters in practice, and in particular the calculation of an adjustment of the high byte (Lines 13–14) is deeply reliant on the original assembly code.

**Listing 4** Python reconstruction of Entombed’s PRNG

```
1  def byte(x):     return x & 0xff
2  def lsb(x):      return byte(x)
3  def msb(x):      return byte(x >> 8)
4  def bit9(x):     return (x >> 8) & 1
5
6  def entombedPRNG(seed):
7          prng = seed
8
9          prng = prng * 5
10         prngH = msb(prng)
11         if prng > 65535:
12                 # "cbit" refers to the 6507's carry bit
13                 cbit = bit9( byte(4 * lsb(seed)) + lsb(seed) )
14                 cbit = bit9( byte(msb(seed << 2) + cbit) + msb(seed) )
15                 prngH = byte( prngH + cbit )
16         prngL = byte( lsb(prng) + 1 )
17         prng = (prngH << 8) + prngL
18
19         return prng
20
21 if __name__ == '__main__':
22         for s in range(256):
23                 seed = (s << 8) | s
24                 for i in range(65536):
```

```
            seed = entombedPRNG(seed)
            print '%04X' % seed
```

When the code in Listing 4 is executed as a program (as opposed to being imported as a module), it cycles through all possible seed values and generates the maximum possible amount of pseudo-random numbers for each seed. Note that Entombed's code duplicates a single byte value in both high and low bytes of the initial 16-bit seed, meaning that 256 initial seed values do in fact cover the entire range of initial seeds.

To verify correctness of the resulting output, we ran Entombed in MAME (MESS) 0.191 as follows, where `script` is given in Listing 5.

```
mess64 —debug —debugscript script —log —window a2600 —cart Entombed.bin
```

The debugger script cajoles the emulator into repeatedly running Entombed's PRNG code through the same sequence as the above Python code, logging each generated value for later comparison. For reference, `$bca5` is the entry point of the PRNG, and `$bcc9` is the PRNG's RTS instruction.

**Listing 5** PRNG verification script

```
bpset bcc9,temp1 < 10000,{ logerror "%02x%02x\n", b@dd, b@de; temp1++; pc=bca5; g }
bpset bcc9,temp1 >= 10000 && temp0 < ff,{ ++temp0; printf "%04X", temp0; b@dd=temp0; b@de=temp0; temp1=0; pc=bca5; g }
bpset bcc9,temp1 >= 10000 && temp0 >= ff
temp0 = temp1 = 0
pc = bca5
w@dd = 0
g
```

## About the authors

**John Aycock** is an associate professor in the Department of Computer Science at the University of Calgary. Contact him at aycock@ucalgary.ca.

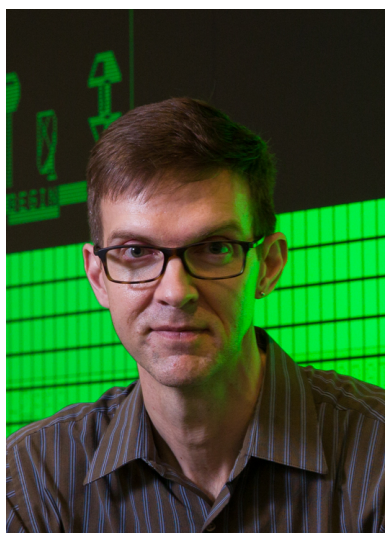

**Tara Copplestone** is a PhD student at the Universities of Aarhus and York. Her work entails the creation, analysis and documentation of videogames for and about archaeology. Contact her at tjc528@york.ac.uk.

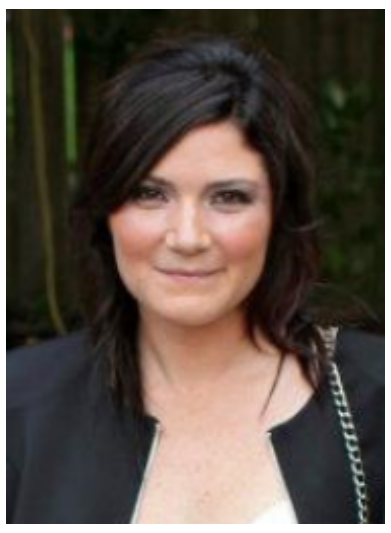